\documentclass[preprint,superscriptaddress,nofootinbib]{revtex4-2}

\usepackage[colorlinks=true, linkcolor=blue, citecolor=blue, urlcolor=blue]{hyperref}
\usepackage{amsmath, graphicx}

\begin{document}

\begin{flushright}
{
OU-HET-1260
}
\end{flushright}

\title{
Super-critical primordial black hole formation \\
via delayed first-order electroweak phase transition
}

\author{Katsuya Hashino}
\email{hashino@fukushima-nct.ac.jp}
\affiliation{
National Institute of Technology, Fukushima College, Iwaki, Fukushima 970-8034, Japan
}

\author{Shinya Kanemura}
\email{kanemu@het.phys.sci.osaka-u.ac.jp} 
\affiliation{
Department of Physics, Osaka University, Toyonaka, Osaka 560-0043, Japan
}

\author{Tomo~Takahashi}
\email{tomot@cc.saga-u.ac.jp} 
\affiliation{
Department of Physics, Saga University, Saga 840-8502, Japan
}

\author{Masanori Tanaka}
\email{tanaka@pku.edu.cn}
\affiliation{
Center for High Energy Physics, Peking University, Beijing 100871, China
}

\author{Chul-Moon Yoo}
\email{yoo.chulmoon.k6@f.mail.nagoya-u.ac.jp}
\affiliation{
Division of Particle and Astrophysical Science, Graduate School of Science, Nagoya University, Nagoya 464-8602, Japan
}

\begin{abstract}

The delay of the first-order electroweak phase transitions (EWPT) may lead to the emergence of baby universes inside wormhole structures due to the large vacuum energy density in false vacuum domains.
Observers outside the false vacuum domains observe them as primordial black holes (PBHs), categorized as super-critical PBHs. 
We specifically investigate the dynamics of PBH formation due to delayed first-order EWPTs by solving the equations of bubble wall dynamics. 
We numerically confirm that such super-critical PBHs can be formed by the delayed first-order EWPT assuming spherically symmetric false vacuum domains with the thin-wall approximation for its boundary.
Our numerical results show that a PBH formation criterion utilizing characteristic timescales is more appropriate than the conventional criterion based on density fluctuations.
Employing our numerical results, we update the parameter regions of new physics models which can be explored by current and future constraints on the PBH abundance.

\end{abstract}

\maketitle
\newpage


\section{Introduction}

Primordial black holes (PBHs) are black holes formed in the early Universe due to the emergence of large density fluctuations~\cite{Hawking:1971ei,Carr:1974nx, Carr:1975qj}. 
Among many models, one of the most interesting scenarios is the PBH formation associated with a vacuum bubble formation~~\cite{Berezin:1983dz,Berezin:1982ur,Blau:1986cw,Tanahashi:2014sma,Garriga:2015fdk,Deng:2016vzb,Deng:2017uwc,Deng:2020mds,Gouttenoire:2023gbn},
from which we may be able to explore models in the early universe predicting the phase transition via PBH observations.

Recently, there has been renewed interest in the possibility that PBHs could be formed by first-order phase transitions at the early universe~\cite{Sato:1981bf,Kodama:1981gu,Kodama:1982sf,Maeda:1981gw,Jedamzik:1999am,Lewicki:2021pgr,Liu:2021svg,Hashino:2021qoq,Kawana:2022olo,Lewicki:2023ioy,Hashino:2022tcs,Gouttenoire:2023naa,Gouttenoire:2023pxh,Banerjee:2023qya,Conaci:2024tlc,Lewicki:2024ghw,Flores:2024lng,Kanemura:2024pae,Cai:2024nln,Goncalves:2024vkj,Arteaga:2024vde,Banerjee:2024cwv}. 
If the PBHs are formed by this mechanism, we may be able to test models with the first-order phase transitions in the early universe by using PBH observations. 
First-order electroweak phase transitions (EWPTs) are important to explain the baryon asymmetry of the Universe (BAU).
If the baryon asymmetry is produced from the initially baryon symmetric Universe, a mechanism satisfying the Sakharov's conditions is required~\cite{Sakharov:1967dj}. 
In electroweak (EW) baryogenesis, which is a promising scenario to explain the BAU, the strongly first-order EWPT is required to satisfy the third Sakharov's condition of departure from thermal equilibrium~\cite{Kuzmin:1985mm}.
However, in the Standard Model (SM) with the discovered 125\,GeV Higgs boson, the EWPT cannot be first-order~\cite{Kajantie:1996mn,DOnofrio:2014rug,DOnofrio:2015gop}, so that an extension of the SM is necessary for successful EW baryogenesis. 
Such a scenario of EW baryogenesis in various extended Higgs models has been discussed~\cite{Turok:1990in,Turok:1990zg,Cline:1995dg,Carena:1997gx,Cline:1997vk,Cline:1998hy,Cline:2000kb,Cline:2000nw,Carena:2000id,Kainulainen:2002th,Carena:2002ss,Fromme:2006cm,Cirigliano:2006wh,Li:2008ez,Chung:2009qs,Cline:2011mm,Liu:2011jh,Cline:2012hg,Chiang:2016vgf,Jiang:2015cwa,Vaskonen:2016yiu,Guo:2016ixx,Fuyuto:2017ewj,Grzadkowski:2018nbc,Modak:2018csw,Cline:2021iff,Kainulainen:2021oqs,Xie:2020wzn,Enomoto:2021dkl,Enomoto:2022rrl,Aoki:2023xnn,Basler:2021kgq,Aoki:2022bkg,Kanemura:2023juv}. 
These models with the strongly first-order EWPT typically predict a large deviation in the triple Higgs boson coupling~\cite{Grojean:2004xa,Kanemura:2004ch} and gravitational waves (GWs) with milli- to deci- Herz frequencies~\cite{Grojean:2006bp, Kakizaki:2015wua,Hashino:2016rvx,Hashino:2018wee}. 
While they can be explored at next generation high energy colliders and future space-based GW interferometers, an earlier experimental method for testing the first-order EWPT is desired. 
If PBHs are formed from the EWPT, the mass of PBHs is about $10^{-5} M_{\odot}$, where $M_{\odot}$ is the solar mass~\cite{Hashino:2021qoq}, and as discussed in Refs.~\cite{Hashino:2021qoq,Hashino:2022tcs} we may be able to test the first-order EWPT by using current and forthcoming microlensing observations such as Subaru Hyper Suprime-Cam (HSC)~\cite{Niikura:2017zjd}, OGLE~\cite{Niikura:2019kqi}, EROS~\cite{EROS-2:2006ryy}, PRIME~\cite{Kondo_2023}, and Roman telescope~\cite{Fardeen:2023euf}. 

First-order phase transitions in the early Universe proceed through the nucleation and expansion of vacuum bubble walls inside which a certain symmetry is broken unless the symmetry restoration scenario is not considered.
If the first-order phase transition is supercooling, it is possible that the symmetry breaking does not occur for a relatively long time in a large region of space.
In the following, we call the isolated symmetry unbroken regions False Vacuum Domains (FVD). 
Also, we refer to first-order phase transitions that leave large FVDs for a relatively long time as delayed first-order phase transitions.
Outside the FVD, radiation energy component is dominant because the symmetry is already broken. 
The energy density inside the FVD is composed of the radiation and vacuum energy components. 
Although the radiation energy density decreases rapidly as the Universe cools, the vacuum energy density remains almost constant.
Then, an inflationary universe can be realized in a local domain.
The investigation of inflationary baby universes in a local domain of our Universe has a long history~\cite{Sato:1981bf,Kodama:1981gu,Kodama:1982sf,Maeda:1981gw,Berezin:1983dz,Berezin:1982ur,Garriga:2015fdk,Deng:2016vzb,Deng:2017uwc,Deng:2020mds}. 
In particular, observers outside the baby Universes observe them as the PBHs~\cite{Garriga:2015fdk,Deng:2016vzb,Deng:2017uwc,Deng:2020mds}, 
which are referred to as ``super-critical PBH" \cite{Garriga:2015fdk}.
We adopt this nomenclature in this paper. 

Since a large difference in energy density between the inside and outside of the FVD can be produced if the symmetry breaking in the FVD is delayed for a relatively long time~\cite{Liu:2021svg}.
A threshold value of the density contrast is often used as the criterion of PBH formation associated with the first-order phase transition.
That is, if the density fluctuation is larger than a certain critical value $\delta_{C}$, the FVD is judged to collapse into a PBH.

As a critical value, $\delta_{C} \sim 0.45$ is often used in the literature\,\cite{Musco:2012au,Harada:2013epa}. 
However, it has originally been proposed for the PBH formation as a result of the adiabatic fluctuations entering the horizon during the radiation-dominated epoch.
It is not trivial whether the condition $\delta > 0.45$ can also be used as a criterion for the PBH formation due to the delayed first-order phase transitions. 

In this paper, we carefully investigate the dynamics of the PBH formation due to the delayed first-order EWPTs in detail.
For simplicity, we consider a spherically symmetric FVD enclosed by the thin vacuum bubble wall. 
Assuming that there is no friction between the wall and the fluid components, we can derive the equation of motion (EoM) for the FVD boundary based on Israel's junction condition~\cite{Israel:1966rt} as shown in Refs.~\cite{Tanahashi:2014sma,Deng:2016vzb}. 
Then, we can investigate its dynamics by solving the EoM, an ordinary differential equation.
It is known that the diverging area radius of the FVD boundary indicates the super-critical PBH formation~\cite{Blau:1986cw,Deng:2016vzb,Deng:2017uwc,Deng:2020mds}.
We confirm that the super-critical PBHs can indeed be formed in scenarios with the delayed first-order EWPTs by solving the EoM.
Based on the comparison between the time scales associated with the vacuum energy and the Hubble parameter at the horizon crossing, we argue that a criterion for PBH formation using the time scales works almost equivalently to or better than the conventional one $\delta > \delta_{C}$. 

The criterion is also important to investigate physics beyond the standard model of particle physics. 
It has been shown that new physics models with delayed first-order EWPTs can be explored by PBH observations in the Standard Model Effective Field Theory (SMEFT)~\cite{Hashino:2021qoq} using the conventional criterion $\delta > \delta_{C}$.  
This was soon extended to the analysis in the nearly-aligned Higgs Effective Field Theory (naHEFT)~\cite{Kanemura:2021fvp, Kanemura:2022txx, Florentino:2024kkf}, in which physics of the first-order EWPT from non-decoupling loop effects of new physics can be better described~\cite{Postma:2020toi, Banta:2022rwg,Kanemura:2022txx}.
In the present paper, we also employ the framework of the naHEFT. 
We discuss parameter regions in the naHEFT that can be explored by PBH observations by using the machinery developed in the subsequent sections. 

We outline the structure of this paper. 
In Sec.~\ref{sec:naHEFT}, we briefly summarize the definition of the naHEFT. 
In Sec.~\ref{sec:PBH}, the PBH formation due to the delayed first-order EWPTs is described. 
We define the equation for the dynamics of the FVD and show its numerical solutions. 
Then, the appropriate PBH formation criterion is discussed there. 
In Sec.~\ref{sec:results}, predictions on properties of the PBHs formed by the EWPTs are discussed. 
In addition, parameter regions explored by current and future PBH observations are shown within the framework of the naHEFT by using our numerical solutions discussed in Sec.~\ref{sec:PBH}. 
In Sec.~\ref{sec:conclusions}, we give discussions and conclusions.


\section{The nearly aligned Higgs effective field theory \label{sec:naHEFT}}

In this section, we introduce the naHEFT. 
In this EFT framework, the Higgs potential at finite temperatures is expressed by~\cite{Kanemura:2021fvp,Kanemura:2022txx}
\begin{align}
V_{\rm naHEFT}(\phi, T) = V_{\rm SM}(\phi, T) + V^{\rm BSM}_{T=0}(\phi) + V_{T}^{\rm BSM}(\phi, T) \,,
\end{align}
where $V_{\rm SM}(\phi, T)$ is the SM contribution at finite temperature, $V^{\rm BSM}_{T=0}(\phi)$ and $V_{T}^{\rm BSM}(\phi, T)$ are the BSM ones at zero and finite temperatures, respectively.
The field $\phi$ represents the order parameter of the SM Higgs field. 
For the SM part, we take account of the contributions from the weak gauge bosons, Higgs boson and top quark. 
For the BSM parts, $V^{\rm BSM}_{T=0}(\phi)$ and $V_{T}^{\rm BSM}(\phi, T)$ are respectively given by 
\begin{align}
&V^{\rm BSM}_{T=0}(\phi) = \frac{\xi}{4} \kappa_{0} \left[ \mathcal{M}^2(\phi) \right]^2 \ln \frac{\mathcal{M}^2(\phi)}{\mu^2} \,, \\
&V^{\rm BSM}_{T}(\phi, T) = 8 \xi T^4 \kappa_{0} \int^{\infty}_{0} dk^2 k^2 \ln \left( 1 - {\rm sign} (\kappa_{0}) \exp \left[ - \sqrt{k^2 + \frac{\mathcal{M}^2(\phi)}{T} } \right] \right) \,,
\end{align}
where $\xi=1/16\pi^2$, $\kappa_{0}$ characterizes the degrees of freedom of new particles, and 
$\mu$ can be determined by the renormalization of the Higgs potential. 
We here assume that the form factor $\mathcal{M}^2(\phi)$ can be expressed by 
\begin{align}
\mathcal{M}^2(\phi) = M^2 + \frac{\kappa_{p}}{2} \phi^2 \,,
\end{align}
where $M^2$ and $\kappa_{p}$ are real parameters. 
Furthermore, we introduce two parameters $\Lambda$ and $r$ as 
\begin{align}
\Lambda^2 = M^2 + \frac{\kappa_{p}}{2} v^2, \quad 
r = \frac{\frac{\kappa_{p}}{2} v^2}{\Lambda^2} = 1 - \frac{M^2}{\Lambda^2} \,,
\end{align}
where $v = 246\,{\rm GeV}$. 
The physical meaning of $\Lambda$ is the degenerate mass of new particles integrated out. 
The parameter $r$, which is called \textit{the non-decouplingness}, parameterizes the appearance of non-decoupling effects. 
When $r \to 0$, we obtain $\Lambda \sim M$, which indicates that the mass of new particles is independent of the Higgs mechanism. 
Therefore, the corrections to low-energy observables from such new particles can be decoupled from the SM predictions when we consider the limit $\Lambda \gg v$. 
On the other hand, when $r \to 1$, new particles acquire their masses due to the Higgs mechanism ($\Lambda \sim \kappa_p v^2/2$). 
In such a case, the non-decoupling effects become important because low-energy observables can include significant radiative corrections from new particles. 
We can describe decoupling and non-decoupling new physics effects by changing the value of the non-decouplingness $r$. 

In the following, we treat the following three quantities as independent free parameters: 
\begin{align}
\kappa_{0}, \quad \Lambda, \quad r \,.
\end{align}
Once the above three quantities are fixed, we can predict important observables such as the triple Higgs boson coupling $hhh$, the spectrum of GWs and the abundance of PBHs. 
In several extended Higgs models, a large deviation from the SM prediction in the $hhh$ coupling is required to realize the strongly first-order EWPT~\cite{Grojean:2004xa,Kanemura:2004ch}. 
The deviation can be expressed by using the effective potential approximation as~\cite{Kanemura:2021fvp,Kanemura:2022txx,Florentino:2024kkf}
\begin{align}
\frac{\Delta \lambda_{hhh}}{\lambda_{hhh}^{\rm SM}} 
\equiv \frac{\lambda_{hhh}^{\rm new} - \lambda_{hhh}^{\rm SM}}{\lambda_{hhh}^{\rm SM}}, \quad \lambda_{hhh}^{\rm new} = \left. \frac{\partial^3 V_{\rm naHEFT}(\phi, T=0)}{\partial \phi^3} \right|_{\phi = v} \,,
\label{eq:hhh}
\end{align}
where $\lambda_{hhh}^{\rm SM}$ is the $hhh$ coupling in the SM. 

In discussing the GW spectra and PBH formation, the vacuum bubble nucleation rate plays an important role.
The bubble nucleation rate per unit volume and time is given by~\cite{Linde:1981zj}
\begin{align}
\Gamma(T) \simeq T^4 \left( \frac{S_{3}(T)}{2 \pi T} \right)^{3/2} \exp \left( - \frac{S_{3}(T)}{T} \right) \,,
\label{eq:GammaT}
\end{align}
where $S_{3}(T)$ is the three-dimensional Euclidean action for the bounce solution. 
By using Eq.\,\eqref{eq:GammaT}, the fraction of the false vacuum region can be expressed by 
\begin{align}
F(t) = \exp \left[ - \frac{4\pi}{3} \int^{t}_{t_{i}} dt' \Gamma(T(t')) a^3(t') \eta^3(t, t')\right] \,, 
\end{align}
where $t_{i}$ is the time when a bubble is nucleated in the Universe. 
The function $T(t)$ implies the temperature determined at the time $t$. 
The factor $\eta(t, t')$, which indicates the time evolution of the comoving radius of vacuum bubbles, is defined by 
\begin{align}
\eta(t , t') \equiv \eta_0(t') + \int^{t}_{t'} d\tilde{t} \frac{v_{w}}{a(\tilde{t})} \,,
\end{align}
where 
$\eta_0(t')$ is the initial critical radius of the vacuum bubbles. 
Since $\eta_0(t')$ is generally much smaller than the Hubble scale, we neglect it in the following analysis.
$v_w$ is the mean wall velocity that is regarded as a free parameter in this paper.
The evolution of radiation energy component in the false vacuum region
can be described by
\begin{align}
\frac{d\rho_{R}}{dt} + 4H \rho_{R} = - \frac{d \rho_{V}}{dt} \,,
\end{align}
with $\rho_{V}(t) = \Delta V(t)$, where $\Delta V(t)$ is the potential height difference between false and true vacua\footnote{
It has been often assumed that the volume average of the vacuum energy density can expressed by $\rho_{V}(t) = F(t) \Delta V(t)$. 
However, since the local temperature in the false vacuum domain would be relevant to the bubble nucleation rate, we do not take the volume average. 
Nevertheless, we have confirmed that the difference in the final result is insignificant.
}.
The Hubble parameter is determined by the Friedmann equation, which is given by 
\begin{align}
\label{eq:Hubble}
H^2 = \left( \frac{\dot{a}}{a}\right)^2= \frac{1}{3 M_{p \ell}^2} (\rho_{V} + \rho_{R}) \,, 
\end{align}
where $M_{p \ell}$ is the reduced Planck mass, and a dot represents a derivative with respect to time. 
When the Hubble parameter is determined by solving Eq.\,\eqref{eq:Hubble}, the relation between the time and temperature can be obtained by
\begin{align}
\frac{dt}{dT} = - \frac{1}{TH(T)} \,.
\end{align}
Using the above equation, we can determine the temperature as a function of time $T(t)$.

Before finishing this section, let us introduce the GW spectrum produced by the EWPT, which can also be used as a probe of EWPT.
Among several sources of GW generated from first-order phase transitions, the contribution of sound waves is dominant, the spectrum of which is given by~\cite{Caprini:2015zlo} 
\begin{align}
  \Omega_{\rm sw}(f)h^2
  = 2.65 \times 10^{-6} v_{w} \left( \frac{H}{\beta} \right) \left( \frac{\kappa_{v} \alpha }{1+\alpha} \right)^2 \left( \frac{100}{g_{*}}\right)^{1/3} (f/f_{\rm sw})^3 \left( \frac{7}{4+3(f/f_{\rm sw})^2} \right)^{7/2},
\end{align} 
where $\alpha$ and $\beta/H$ are the latent heat normalized by radiative energy density and the inverse of the duration of phase transition at the nucleation temperature $T_n$, respectively.
The definition of these quantities is the same as that given in Ref.,\cite{Grojean:2006bp}.
The value of effective degrees of freedom $g_{*}$ is based on the result in Ref.\,\cite{Husdal:2016haj}. 
The factor $\kappa_v$ is the efficiency factor~\cite{Espinosa:2010hh}. 
The peak frequency $f_{\rm sw}$ is given by~\cite{Caprini:2015zlo}
\begin{align}
  f_{\rm sw} = 1.9 \times 10^{-2} {\rm mHz} \frac{1}{v_{w}} \left( \frac{\beta}{H}  \right) \left( \frac{T_{n}}{100 {\rm GeV}} \right) \left( \frac{g_{*}}{100}\right)^{1/6}.
\end{align}
The detectability of the GW spectra in the naHEFT has been discussed in Ref.~\cite{Kanemura:2022txx}.


\section{FVD boundary dynamics and PBH formation \label{sec:PBH}}

\subsection{Equations of motion and initial conditions}

The FVDs are separated by the vacuum bubble walls from the symmetry broken region. 
Since the energy density in the FVD is isotropic and homogeneous, it is expected that the metric inside the FVD takes the Friedmann-Lema\^itre-Robertson-Walker (FLRW) metric. 
For the outside of the FVD, its metric is set to be spherically symmetric because the FVD is assumed to be spherically symmetric.
If the thin-wall approximation can be applied to the FVD boundary, we can employ Israel's junction conditions~\cite{Israel:1966rt}, which are derived by integrating the Einstein equations in the vicinity of the
thin wall.
Through Israel's junction conditions, the energy-momentum tensor of the thin wall causes the discontinuity of the extrinsic curvature on both sides of the wall.
Using Israel's junction conditions, we can obtain the EoM for the FVD boundary as given in Eq.~\eqref{eq:EOM} below, whose derivation is explained in Appendix~\ref{sec:EoM}.
The EoM is expressed in terms of the comoving area radius of the FVD boundary $\chi$ as~\cite{Deng:2020mds}
\begin{align}
\label{eq:EOM}
\ddot{\chi} + (4 - 3a^2 \dot{\chi}^2) H \dot{\chi} + \frac{2}{a^2 \chi} (1 - a^2 \dot{\chi}^2) = \left( \frac{3\sigma}{4M_{p\ell}^2} - \frac{\rho_{V}}{\sigma}  \right) \frac{(1 - a^2 \dot{\chi}^2)^{3/2}}{a} \,,
\end{align}
where $a$, $H$ and $\rho_{V}$ are the scale factor, Hubble parameter and the vacuum energy in the FVD, respectively. 
Since we assume that the interior of the FVD is described by the FLRW universe filled with radiation and the vacuum energy whose energy densities are denoted as $\rho_{\rm R}$ and $\rho_V$, respectively. 
Then, the scale factor $a$ and the Hubble expansion rate $H$ satisfy Eq.~\eqref{eq:Hubble}.
$\sigma$ is the surface energy of the FVD boundary, which is approximately given by~\cite{Linde:1981zj}
\begin{align}
\sigma \simeq \int_{0}^{\infty} dr \left[ \frac{1}{2} \left( \frac{d\phi_{B}}{dr} \right)^2 + V_{\rm eff}(\phi_{B}, T(t_{\rm in})) \right] \,,
\label{eq:surface}
\end{align}
where $\phi_{B}(r)$ and $V_{\rm eff}(\phi, T)$ are a bounce solution and a finite temperature potential at the temperature $T$, respectively. 
We evaluate the value of $\sigma$ at the initial time $t=t_{\rm in}$ and consider $\sigma$ as a constant parameter for the FVD boundary dynamics for simplicity. 
In the following, we take $V_{\rm eff}(\phi, T) = V_{\rm naHEFT}(\phi, T)$. 
We utilize \texttt{cosmoTransitions}~\cite{Wainwright:2011kj} to obtain the bounce solution numerically.

For simplicity, the comoving FVD boundary is assumed to be at rest at the initial time $t=t_{\rm in}$.
That is, we assume $\dot \chi(t_{\rm in})=0$. We set the initial time $t_{\rm in}$ to the percolation time when the fraction of the false vacuum domain outside the FVD satisfies $F(t)=0.7$~\cite{Enqvist:1991xw}. 
Then, we can define the comoving horizon radius $\chi_{\rm h}$ as
\begin{align}
\label{eq:EOM_boundary}
a(t_{\rm in}) \, \chi_{\rm h} = \frac{M_{\rm MS}(t_{\rm in})}{4 \pi M_{p \ell}^2} \,,
\end{align}
where $M_{\rm MS}$ is the Misner-Sharp mass defined just outside the FVD~\cite{Misner:1964je}, which is given by~\cite{Deng:2020mds,Yoo:2022mzl}
\begin{align}
\label{eq:Misner_Sharp}
M_{\rm MS}(t) = \frac{4\pi}{3} (\rho_{V}(t) + \rho_{\rm R}(t)) R(t)^3 + 4 \pi \sigma \gamma R(t)^2 + 4 \pi \sigma H(t) \sqrt{\gamma^2 - 1} R(t)^3 - \frac{\pi \sigma^2}{M_{p \ell}^2} R(t)^3 \,,
\end{align}
with $R(t) = a(t) \chi(t)$ and $\gamma = 1/\sqrt{1 - a^2 \dot{\chi}^2}$. 
In the analyses, we consider several values of the initial shell radius comparable to the horizon radius $\chi_{\rm h}$. 
It should be noted that the wall velocity $v_w$ and $\dot \chi$ are independent variables. 
As defined above, $\chi$ describes the time evolution of the area radius of a FVD that may form a PBH.
On the other hand, in our analyses, $v_w$ is the mean value of the wall velocity, which is expected to be close to the light speed if the phase transition is supercooling~\cite{Lewicki:2021pgr}.

\subsection{Sub/super-critical solution and the analytic criteria}

Once the solution of Eq.\,\eqref{eq:EOM} is obtained, the time evolution of the FVD boundary is determined. 
In general, two kinds of solution can be obtained~\cite{Garriga:2015fdk,Deng:2016vzb,Deng:2017uwc,Deng:2020mds}:
\begin{enumerate}
\item Super-critical: the radius of the 
FVD boundary
never reaches zero 
\item Sub-critical: the radius of the 
 FVD boundary 
becomes zero in a finite time
\end{enumerate}
For the super-critical case, the radius of the FVD boundary continues to increase following the accelerated expansion inside the wall, and a baby universe is realized in the local region inside the wall~\cite{Berezin:1982ur, Berezin:1983dz}. 
The baby universe is connected to our Universe through an untraversable wormhole throat like the Einstein-Rosen bridge for the Schwarzschild spacetime, and it is regarded as a PBH from outside observers.
Such PBHs are called the super-critical PBHs~\cite{Garriga:2015fdk}.

We note that, in the sub-critical case, the realization of the PBH formation is rather non-trivial and two possible PBH formation processes can be considered.
One is the direct PBH formation due to the collapse of the spherical domain wall. 
However, in this case, the wall width, which is neglected in the thin wall approximation, must be smaller than the Schwarzschild radius for the energy carried by the FVD boundary. 
The width of the boundary is expected to be given by $\sim 1/E_{\rm EW}$ with $E_{\rm EW}$ being the EW energy scale. 
The initial total energy of the FVD boundary $\rho_{\rm in}$ can be estimated as $ \rho_{\rm in} \sim \sigma/H^2(t_{\rm in})\sim M_{pl}^2/E_{\rm EW}$, where we have used $\sigma\sim E_{\rm EW}^3$ and $H(t_{\rm in})\sim E_{\rm EW}^2/M_{pl}$. 
If we assume the total energy is almost conserved\footnote{
The initial mass of the black hole formed through the sub-critical collapse in the dust universe is found to be $\sim 17\sigma/H_{\rm hc}^2$ with $H_{\rm hc}$ being the Hubble expansion rate at the horizon crossing time under the thin wall approximation~\cite{Tanahashi:2014sma}.}, the Schwarzschild radius for the total mass is given by $\sim 1/E_{\rm EW}$, which is of the same order as the boundary width.
Therefore, in order to achieve the PBH formation in the case of the sub-critical collapse, the FVD boundary must gain sufficient kinetic energy through its shrinking dynamics.
In general, the vacuum energy inside the FVD can be converted into the kinetic energy of its boundary.
However, it is expected that the energy stored inside the FVD is also converted into other components such as the kinetic energy of plasma, particle masses, and the energy of GWs. 
Thus, it is not trivial to determine how much energy is finally carried by the FVD boundary. 
The other possible sub-critical PBH formation process is the PBH formation due to the subsequent accretion of the radiation. Even if PBHs cannot directly be formed due to the collapse of the FVD, the radiation that passes through its boundary gains inward velocity due to the Ricci focusing effect in terms of the Raychaudhuri equation as shown in Ref.~\cite{Tanahashi:2014sma} for dust cases. 
Then PBHs may be formed as a result of the subsequent radiation fluid dynamics.
However, to investigate this dynamics, we need a non-linear relativistic simulation. 
These are left for future work, and here we focus on the PBH formation through the super-critical domain wall dynamics in this paper.

Now let us consider the feasibility of super-critical PBH formation by comparing some time scales discussed in Ref.\,\cite{Garriga:2015fdk}. 
One of the relevant time scales is the Hubble time at the horizon crossing. 
Assuming that the background universe is described by the radiation-dominated one, we obtain 
\begin{align}
\label{eq:tH}
t_H \equiv \frac{1}{2H_{\rm hc}(\chi_{\rm in})} =\frac{1}{2}H_{\rm in}R_{\rm in}^2 \,,
\end{align}
where we used $1/H_{\rm hc}(\chi_{\rm in})=a(t_{\rm hc}(\chi_{\rm in}))\chi_{\rm in}$. 
The expression on the right-hand side can be evaluated at the initial time irrespective of the subsequent dynamics of the bubble wall. 
We adopt this expression as the definition of $t_H$. 
We also define the time scale after which the vacuum energy dominates the expansion of the universe inside the FVD as
\begin{align}
t_{V} = \frac{1}{2}\sqrt{\frac{3M_{p \ell}^2}{\rho_{V}}} \,.
\end{align}
Another relevant time scale is the one associated with the surface energy density $\sigma$ defined by
\begin{align}
t_{\sigma} = \frac{1}{2 \pi G \sigma} = \frac{4 M_{p \ell}^2}{\sigma} \,.
\end{align}
In our settings, the time scale $t_{\sigma}$ is much larger than $t_{H}$ and $t_{V}$. 
It means that the surface energy contribution to PBH formation is negligible, which can be understood by noticing that
\begin{align}
\frac{t_{V}}{t_{\sigma}} \sim \frac{\sqrt{3}\sigma}{ 8 M_{p \ell}\sqrt{\rho_{V}}} \sim \frac{\sqrt{3} v}{8 M_{p\ell}} \ll 1 \,.
\end{align}
Our finding is consistent with the result in Ref.~\cite{Flores:2024lng}. 
Then, we expect that the condition for the super-critical PBH formation is given by~\cite{Garriga:2015fdk} 
\begin{align}
\label{eq:tvth}
t_V\lesssim t_H. 
\end{align}
The validity of this expectation will be verified numerically in the next section. 

Let us introduce another criterion for PBH formation based on the density fluctuation, which has been widely adopted in the literature. 
The density fluctuation $\delta$ around the FVD is defined by 
\begin{align}
\delta = \frac{ \rho_{\rm \,in}(t_H) - \rho_{\rm out}(t_H)}{ \rho_{\rm out}(t_H)} \,,
\end{align}
where $\rho_{\rm \, in}(t_H)$ and $\rho_{\rm out}(t_H)$ are the total energy density inside and outside of the FVD at $t=t_H$ defined by Eq.~\eqref{eq:tH}, respectively. 
Conventionally, when the maximum value of $\delta$ satisfies $\delta_{\rm max} > \delta_{C}$, it is concluded that PBHs can be formed~\cite{Liu:2021svg}. 
However, it should be noted that the value of the density fluctuation depends on the time at which it is evaluated. 

In the current setting, the background dynamics is described by the radiation-dominated universe at the early stage and then by the vacuum energy-dominated universe at some later time.
Since $t_H$ is defined by assuming the radiation-dominated background (see Eq.~\eqref{eq:tH}), $t_H$ can differ from the actual horizon crossing time specified by $1/H=M_{\rm MS}/(4\pi M_{pl}^2)$. 
We note that, although evaluating $\delta$ at $t=t_H$ seems to be natural and works well, the use of some other time involving the one relevant to the vacuum energy might also be possible, which may improve the estimate.
Moreover, the FVD boundary may not experience the horizon crossing if the initial radius is sufficiently large. 
Therefore, we adopt the criterion that the PBH formation occurs when the super-critical solution of the EoM in Eq.\,\eqref{eq:EOM} is obtained.

\subsection{Numerical results for the FVD boundary dynamics}

In this section, we present our results obtained from numerical analyses based on the discussion in the previous subsections.

\begin{figure}[t]
\includegraphics[width=0.49\textwidth]{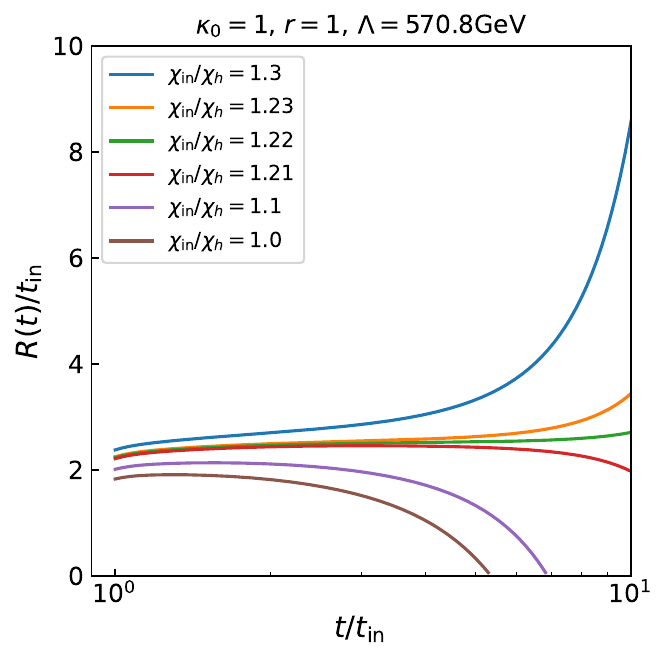}
\includegraphics[width=0.49\textwidth]{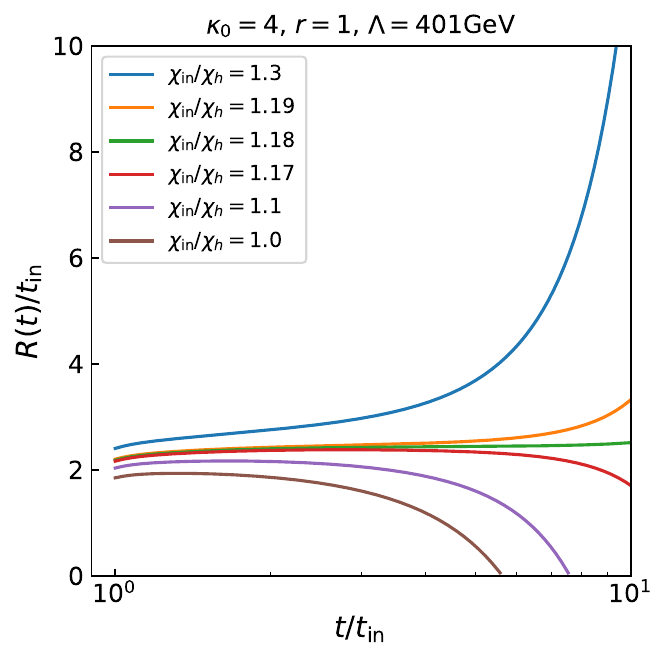}
\caption{
Time evolution of the FVD boundary in the naHEFT. 
We take two benchmark points with $(\kappa_{0},\, r, \, \Lambda)= (1, \, 1, \, 570.8\,{\rm GeV})$ and $(4, \, 1, \, 401\,{\rm GeV})$ in left and right panels, respectively.
We here assume that the wall velocity is $v_w=0.7$. 
As the value of $\chi_{\rm in}$ becomes larger, the super-critical solutions can be obtained in which the radius of the FVD boundary increases monotonically. 
\label{fig:rbubble}
}
\end{figure}

In Fig.\,\ref{fig:rbubble}, the time evolution of the FVD boundary is shown for various values of $\chi_{\rm in}$ in the naHEFT with $(\kappa_{0}, \, r, \, \Lambda, v_{w}) = (1, \, 1, \, 570.8\,{\rm GeV}, \, 0.7)$ and $(4, \, 1,\, 401\,{\rm GeV}, \, 0.7)$.
These solutions are obtained by solving the EoM in Eq.\,\eqref{eq:EOM} with the boundary conditions~\eqref{eq:EOM_boundary}. 
As the value of $\chi_{\rm in}$ increases, the radius of the FVD boundary increases monotonically. 
This implies that the super-critical solution can be realized when $\chi_{\rm in}$ is relatively large as is expected through the discussions given in the previous subsections.
In contrast, sub-critical solutions can be obtained when $\chi_{\rm in}$ is relatively small. 
In the following, we conclude that super-critical PBHs can be formed if a solution of the EoM in Eq.\,\eqref{eq:EOM} satisfies the conditions $R(t_{\rm end})>R(t_{\rm in})$ and $\dot{R}(t_{\rm end})>0$, where the time $t_{\rm end}$ is defined by $T(t_{\rm end})=0.1\,{\rm GeV}$.

\begin{figure}[t]
\includegraphics[width=0.49\textwidth]{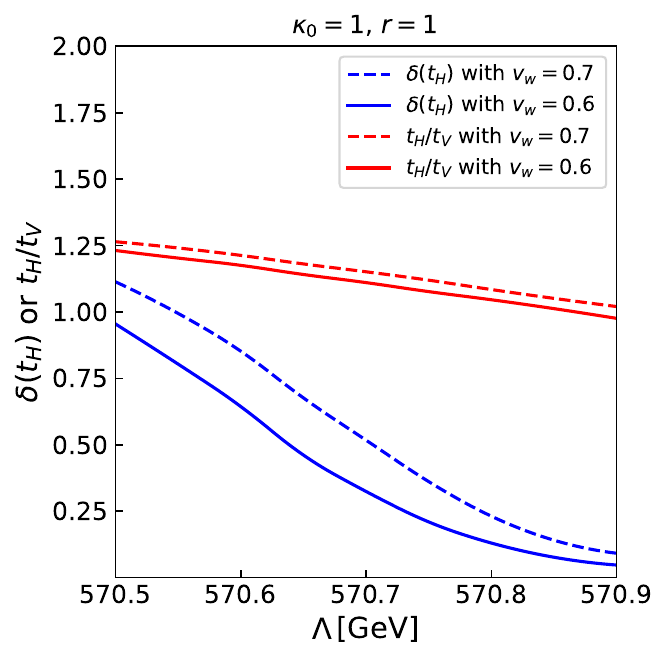}
\includegraphics[width=0.49\textwidth]{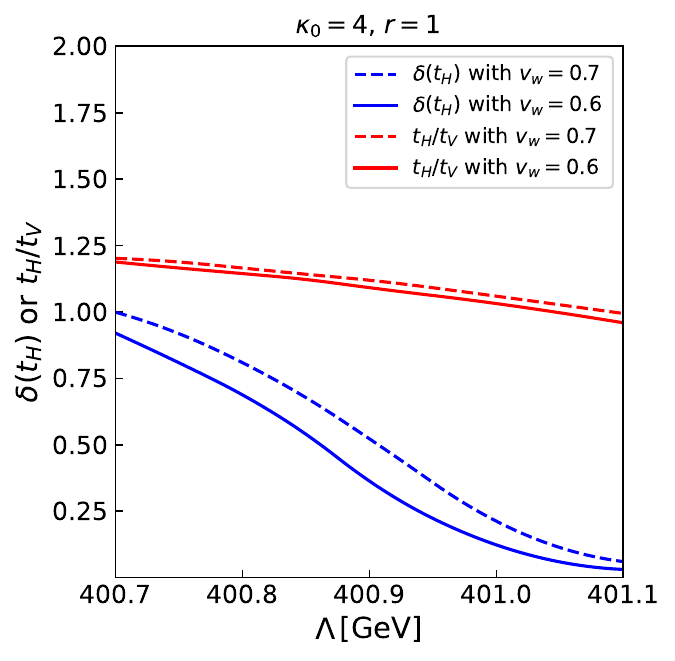}
\caption{
Density fluctuation $\delta$ at $t = t_H$ and the time ratio $t_{H}/t_{V}$ in the naHEFT with $(\kappa_{0},\, r)= (1, \, 1)$ and $(4, \, 1)$. 
The blue solid and dashed lines are the prediction on $\delta(t_{H})$ with $v_{w} = 0.6\,(0.7)$. 
The red solid and dashed line shows the requirement of the time ratio $t_{H}/t_{V}$ to realize the super-critical PBH formation. 
\label{fig:delta_Tin}
}
\end{figure}

In Fig.~\ref{fig:delta_Tin}, $\Lambda$ and $v_w$ dependencies of the density fluctuation $\delta$ at $t = t_H$ and the time ratio $t_{H}/t_{V}$ are shown in the naHEFT with $(\kappa_{0},\, r)= (1, \, 1)$ and $(4, \, 1)$. 
The $v_w$ dependence appears only through the value of $t_{\rm in}$ which we define such that $F(t_{\rm in})=0.7$.
The red solid and dashed lines indicate the requirement of the time ratio $t_{H}/t_{V}$ for the super-critical PBH formation for each wall velocity.
The blue solid and dashed lines show the predictions on $\delta(t_{H})$ with $v_{w}= 0.6 \,(0.7)$.
Fig.\,\ref{fig:delta_Tin} shows that the value of $t_H/t_V$ slightly depends on the model parameters in the naHEFT and $v_{w}$ but the dependencies are much weaker than those for $\delta(t_H)$.
Therefore, we can conclude that Eq.\,\eqref{eq:tvth} is more appropriate for the PBH formation criterion than that in terms of $\delta(t_H)$ as suggested in Ref.\,\cite{Garriga:2015fdk}. 
Thus, if one is eager to confirm the feasibility of the super-critical PBH formation without solving the EoM, it is appropriate to use the criterion in Eq.\,\eqref{eq:tvth} instead of the conventional criterion $\delta_{\rm max}>\delta_{C}$.


\section{PBH abundance and constraints on model parameters \label{sec:results}}

When the super-critical solution in Eq.~\eqref{eq:EOM} is obtained, the properties of PBHs, such as the mass and the fraction of PBHs, can be estimated. 
We can evaluate the PBH mass by using the Misner-Sharp mass, given in Eq.~\eqref{eq:Misner_Sharp}, as 
\begin{align}
\label{eq:PBHmass}
M_{\rm PBH} \equiv M_{\rm MS}(t_{\rm PBH}) \,,
\end{align}
where $t_{\rm PBH}$ is the PBH formation time.
For the case in which the super-critical PBH can be formed, we regard $t_{\rm in}$ as the typical PBH formation time. 
In most cases of our interest, the first two terms in the Misner-Sharp mass in Eq.\,\eqref{eq:Misner_Sharp} give dominant contributions compared to the others. 
In this case, we obtain $R(t_{\rm PBH}) \sim H^{-1}(t_{\rm PBH})$ and the PBH mass can be given by
\begin{align}
M_{\rm PBH} \sim 4 \pi M_{p \ell}^2 H(t_{\rm PBH})^2 R(t_{\rm PBH})^3 \sim 4 \pi M_{p \ell}^2 H^{-1}(t_{\rm PBH}) \,.
\end{align}
This estimate is consistent with the conventional evaluation of the PBH mass~\cite{Hashino:2021qoq,Hashino:2022tcs}.

The fraction of PBHs in dark matter density can be determined by the probability that the first-order phase transition does not occur in a spherically symmetric FVD by a certain time.
For the super-critical dynamics to be completed, the false vacuum domain must be kept until the vacuum energy dominates in the FVD. 
Therefore, we require that the first-order phase transition does not occur in the FVD until $t=t_{V}$. 
The probability for a given value of $\chi_{\rm in}$ is given by 
\begin{align}
P(\chi_{\rm in}) = \exp \left[ - \frac{4\pi}{3} \int_{t_{i}}^{t_V} dt \frac{a^3(t)}{a^3(t_{\rm in})} R(t_{\rm in})^3 \Gamma(T(t)) \right]\,.
\end{align}
Then, the PBH fraction $f_{\rm PBH}$ can be estimated by~\cite{Hashino:2021qoq,Hashino:2022tcs}
\begin{align}
\label{eq:fpbh}
f_{\rm PBH} \sim 1.49 \times 10^{11} \left( \frac{0.25}{\Omega_{\rm DM} }\right) \left( \frac{T(t_{\rm PBH})}{100\,{\rm GeV}}\right) P(\chi_{\rm th}) \,,
\end{align}
where $\Omega_{\rm DM}$ is the fraction of energy density of dark matter, and $\chi_{\rm th}$ is the threshold value of $\chi_{\rm in}$ for the realization of super-critical PBH formation. 

\begin{figure}
\centering
\includegraphics[width=0.49\textwidth]{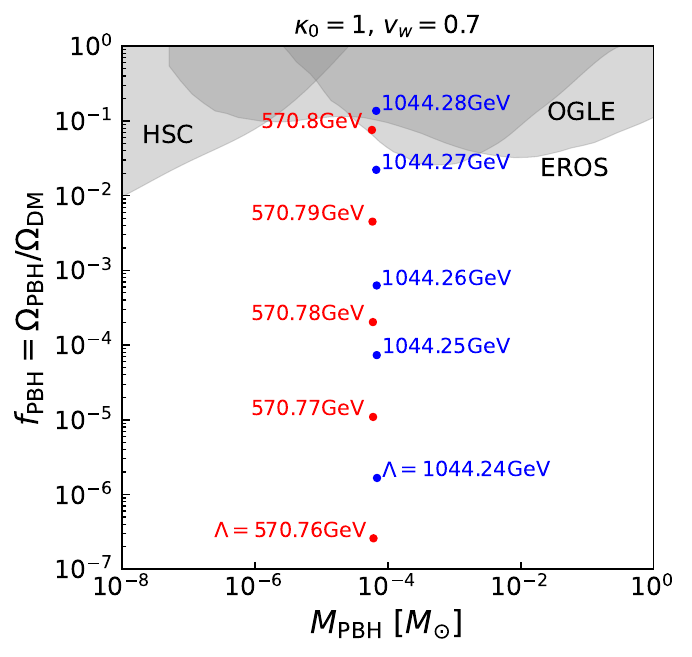}
\includegraphics[width=0.49\textwidth]{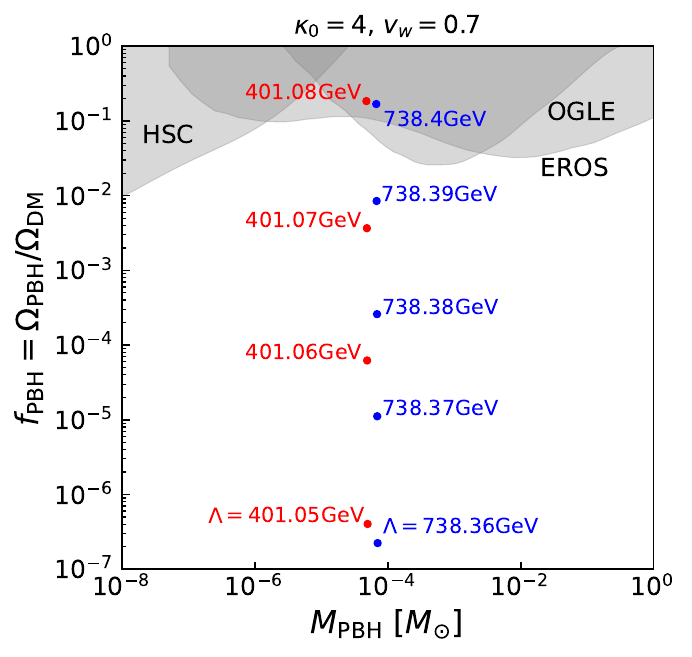}
\caption{
The prediction on the fraction and mass of PBHs in the naHEFT with $\kappa_{0} =1$ and $\kappa_{0} = 4$. 
The wall velocity is assumed as $v_w=0.7$ for both cases. 
The red (blue) points are obtained by solving the EoM with $r = 1$ ($r=0.5$) for each $\Lambda$. 
The gray shaded regions show the current constraints from Subaru HSC~\cite{Niikura:2017zjd}, OGLE~\cite{Niikura:2019kqi} and EROS~\cite{EROS-2:2006ryy}. 
\label{fig:fpbh_mpbh}
}
\end{figure}

\begin{figure}
\centering
\includegraphics[width=0.49\textwidth]{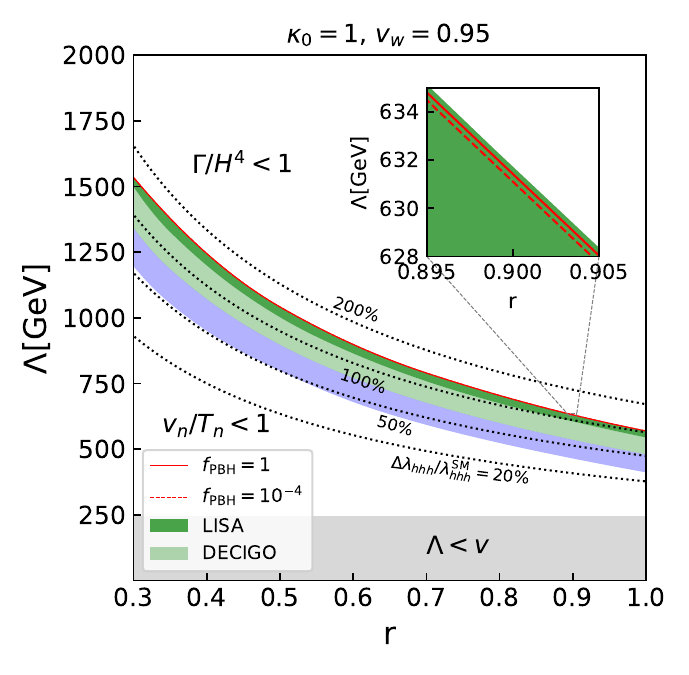}
\includegraphics[width=0.49\textwidth]{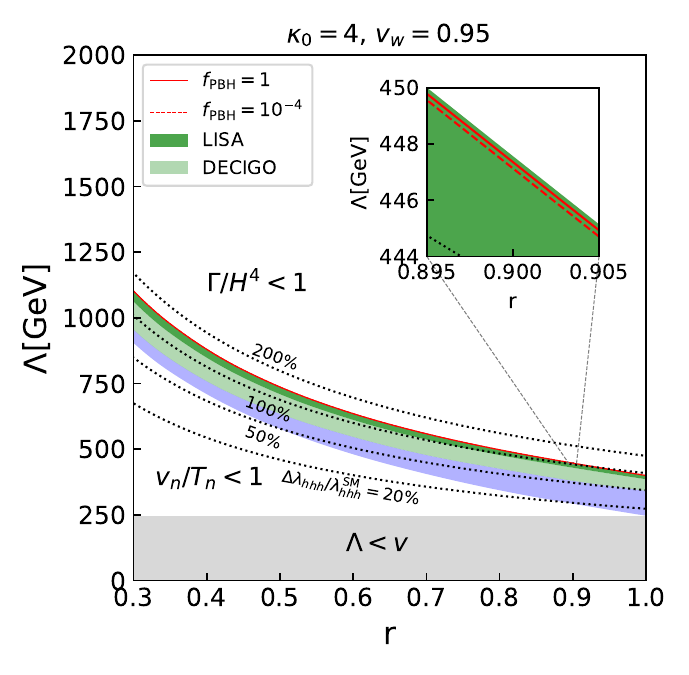}
\caption{
Parameter regions explored by $hhh$ measurements, GW, and PBH observations in the naHEFT with $(\kappa_{0},\, v_{w})= (1, 0.95)$ and $(4, \, 0.95)$.  
The red solid and dashed lines indicate the contours of the PBH fraction with $f_{\rm PBH} = 1$ and $f_{\rm PBH} = 10^{-4}$, respectively. 
The parameter region between these red contours may be explored by future microlensing observations such as PRIME~\cite{Kondo_2023} and Roman telescope~\cite{Fardeen:2023euf}. 
The green and dark green regions may be explored by DECIGO~\cite{Yagi:2011wg} and LISA~\cite{Klein:2015hvg}, respectively. 
The blue region can be tested by using future $hhh$ coupling measurements such as in High-Luminosity LHC (HL-LHC)~\cite{Cepeda:2019klc} and International Linear Collider (ILC)~\cite{Bambade:2019fyw}. 
In the gray region, the requirement of the naHEFT ($\Lambda < v$) is not satisfied. 
The small panel in the upper right corner shows the region around $0.895 < r < 0.905$. 
\label{fig:fpbh}
}
\end{figure}

In Fig.\,\ref{fig:fpbh_mpbh}, it is shown that the prediction on the mass and fraction of PBH in the naHEFT with $\kappa_{0}=1$ and $\kappa_{0} = 4$. 
The wall velocity is assumed to be $v_w=0.7$ for both cases. 
The red (blue) points are obtained by solving the EoM\,\eqref{eq:EOM} with $r=1$ ($r=0.5$). 
The gray shaded regions show the current constraints from Subaru HSC~\cite{Niikura:2017zjd}, OGLE~\cite{Niikura:2019kqi} and EROS~\cite{EROS-2:2006ryy}. 
The typical super-critical PBH mass formed by the EWPT is $M_{\rm PBH} \sim 4 \times 10^{-5} M_{\odot}$. 
We note that the mass of super-critical PBHs is not sensitive to the parameters in the naHEFT because its mass is determined by the Hubble mass when the EWPT occurs. 
However, the PBH fraction is sensitive to the model parameters. 
This sensitivity comes from the factor $e^{\Gamma(T)} \propto e^{e^{-S_{3}/T}}$ in Eq.\,\eqref{eq:fpbh}. 

In Fig.\,\ref{fig:fpbh}, the parameter region explored by PBH observation is shown. 
In the blue, green, and dark green regions, strongly first-order EWPT can be realized. 
As a condition of strongly first-order EWPT, we use $v_{n}/T_{n}>1$, where $v_{n}$ is the vacuum expectation value at the nucleation temperature $T_{n}$. 
The solid and dashed red lines are the contours of the PBH fraction with $f_{\rm PBH} = 1$ and $f_{\rm PBH} = 10^{-4}$, respectively. 
These contours have been obtained by testing the super-criticality of the FVD boundary. 
The green and dark green regions can be explored by GW observations of DECIGO~\cite{Yagi:2011wg} and LISA~\cite{Klein:2015hvg}, respectively. 
The analyses for the GW spectrum are
performed in the same way as in Refs.~\cite{Kanemura:2022txx,Hashino:2022tcs,Florentino:2024kkf}.
Although the blue region cannot be tested using PBH and GW observations, it can be explored by measuring various Higgs couplings~\cite{Florentino:2024kkf} such as in High-Luminosity LHC (HL-LHC)~\cite{Cepeda:2019klc} and International Linear Collider (ILC)~\cite{Bambade:2019fyw}. 
As an example, the contours of the $hhh$ coupling determined by Eq.\,\eqref{eq:hhh} are shown as black dotted lines. 
In the upper right white region, the completion condition of the phase transition cannot be satisfied~\cite{Turner:1992tz}. 
The gray region corresponds to the one in which the requirement for the validity of the naHEFT $\Lambda > v$ is not satisfied. 
We note that the predictions for the parameter regions explored by the PBH observations are almost the same as those obtained in Ref.\,\cite{Hashino:2022tcs}.
The reason is that the PBH fraction $f_{\rm PBH}$ is sensitive to the model parameters as shown in Fig.\,\ref{fig:fpbh_mpbh}.
As a result, differences in the PBH properties due to different PBH formation criteria can be absorbed by slightly changing the parameters of the naHEFT. 
That is, our improved criterion obtained by solving the EoM in Eq.\,\eqref{eq:EOM} does not drastically change the parameter regions that can be explored by PBH observations.


\section{Discussions and conclusions \label{sec:conclusions}}

We have investigated the PBH formation due to delayed first-order EWPTs. 
It has been numerically confirmed that the super-critical PBHs can indeed be formed by the delayed first-order EWPTs. 
In addition, we have also discussed the criterion for super-critical PBH formation, which can be evaluated without solving the EoM for the FVD boundary in Eq.,\eqref{eq:EOM}. 
We have shown that the criterion in terms of $t_{H}/t_{V}$ is more appropriate than the conventional criterion for the PBH formation $\delta>\delta_{C}$ in the sense that the criterion with $t_{H}/t_{V}$ is relatively insensitive to the model parameters. 
Although the importance of $t_{H}/t_{V}$ has been emphasized in the literature~\cite{Garriga:2015fdk,Deng:2016vzb,Deng:2017uwc,Deng:2020mds}, this work is the first to quantitatively confirm its validity within the framework of the EWPT. 
Moreover, we have investigated the predictions for the parameter region which can be explored by the PBH observations by checking the super-criticality of the FVD boundary in the naHEFT framework. 
As confirmed in Fig.\,\ref{fig:fpbh}, the parameter region we obtained is almost the same as the one in Ref.\,\cite{Hashino:2022tcs}. 
Therefore, our investigation gives further support for the validity of the results given in Ref.\,\cite{Hashino:2022tcs} by explicitly showing the super/sub-critical dynamics of FVD boundary.

In solving the EoM for the FVD boundary in Eq.\,\eqref{eq:EOM}, we have assumed that (i) the FVD is spherically symmetric, and (ii) the thin-wall approximation can be applicable to the FVD boundary. 
In particular, the formation of PBHs may be more difficult when the first assumption is not satisfied\,\cite{Yoo:2020lmg,Escriva:2024lmm}.
To improve these two points, we have to numerically solve Einstein equation for the FVD without the simplification based on the spherical symmetry and Israel's junction conditions, which is left for future work.


\begin{acknowledgments}

This work was supported by JSPS KAKENHI Grant Numbers 20H00160~(SK), 23K17691~(SK, TT), Grant-in-Aid for Research Activity Start-up 23K19052~(KH), and MEXT KAKENHI 23H04515~(TT), 20H05850~(CY), 20H05853~(CY) and 24K07027~(CY). 

\end{acknowledgments}

\appendix 

\section{Equation of motion for the false vacuum domain boundary \label{sec:EoM}}

In this appendix, we describe how to derive the EoM in Eq.\,\eqref{eq:EOM}. 
The calculations in this appendix are based on the formalism employed in Refs.~\cite{Tanahashi:2014sma,Deng:2017uwc,Deng:2020mds}. 

We here assume that the FVD possesses spherical symmetry and that the thin-wall approximation is applicable to the FVD boundary. 
When the dynamics of the FVD boundary depends mainly on the gravitational interactions, its dynamics can be determined by matching the metric inside and outside the FVD. 
Since the energy density inside the FVD is isotropic and homogeneous, it is expected that the metric inside the FVD is described by the FLRW metric
\begin{align}
\label{eq:metric_inside}
ds^2 = dt^2 - a(t)^2(dr^2 + r^2 d\Omega) \,.
\end{align}
For the outside of the FVD, the spherically symmetric metric is realized: 
\begin{align}
\label{eq:metric_outside}
ds^2 = dt^2 - a_{1}(t, r)^2 dr^2 - a_{2}(t, r)^2 r^2 d \Omega \,,
\end{align}
When the thin-wall approximation can be applied to the FVD boundary, Israel's junction conditions can be utilized~\cite{Israel:1966rt}.
The scale factors $a_1(t,r)$ and $a_2(t,r)$ in the metric~\eqref{eq:metric_outside} can be related to the scale factor $a(t)$ in Eq.~\eqref{eq:metric_inside} via Israel's junction conditions as we will see below.
Here we describe the trajectory for the FVD boundary as $(t(\tau), \chi(\tau))$, where $\tau$ and $\chi$ are the proper time and the comoving area radius for the FVD boundary, respectively.
The tangent vector $v^{\mu}$ in the radial direction for the FVD boundary is defined as
\begin{align}
v^{\mu} = (t_{,\tau}, \, \chi_{,\tau}, 0, \, 0) \,,
\end{align}
with $_{,\tau} \equiv d/d\tau$. 
Assuming $t_{,\tau}$ is positive, the normalization condition $v_{\mu} v^{\mu} = -1$ gives 
\begin{align}
t_{,\tau} = \sqrt{1 + a_{1}^2 \chi^2_{, \tau}} \,.
\end{align}
Then, the normal vector on the FVD boundary $\xi_{\mu}$ satisfying $\xi_{\mu} \xi^{\mu} =1$ and $\xi_{\mu} v^{\mu} = 0$ is defined as 
\begin{align}
&\xi^{\mu} = (a_{1} \chi_{,\tau}, \, t_{,\tau}/a_{1}, \, 0, \,0) \,, \\
&\xi_{\mu} = (-a_{1} \chi_{,\tau}, \, a_{1} t_{,\tau}, \, 0, \,0) \,. 
\end{align}
Using $\xi_{\mu}$, the extrinsic curvature $K_{\mu \nu}$ is given by 
\begin{align}
K_{\mu \nu} = h^{\alpha}_{\mu} \nabla_{\alpha} \xi_{\nu} \,,
\end{align}
where $h_{\mu \nu}$ and $\nabla_{\alpha}$ are the induced metric on the FVD boundary and its covariant derivative, respectively.
The induced metric is expressed in terms of $\xi_{\mu}$ as 
\begin{align}
h_{\mu \nu} = g_{\mu \nu} - \xi_{\mu} \xi_{\nu} \,,
\end{align}
where $g_{\mu \nu}$ is the metric for the 4 dimensional spacetime. 

We here introduce the following notation for a physical local quantity $Q$
\begin{align}
[Q]_{0} = Q_{\rm out} - Q_{\rm in}, 
\quad 
\{ Q \}_{0} = Q_{\rm out} + Q_{\rm in}, 
\quad
\bar{Q} = \frac{Q_{\rm out} + Q_{\rm in}}{2} \,,
\end{align}
where ``out" and ``in" mean the physical quantity $Q$ evaluated outside and inside the FVD, respectively. 

The first and second Israel's junction conditions can be expressed as~\cite{Israel:1966rt} 
\begin{align}
\label{eq:1st_Israel}
&[h_{\mu \nu}]_{0} = 0 \,, \\
&[K_{\mu \nu}]_{0} = \frac{1}{M_{p \ell}^2} \left( - S_{\mu \nu} + S \frac{h_{\mu \nu}}{2} \right) \,,
\label{eq:2nd_Israel}
\end{align}
where $S_{\mu \nu}$ is the energy-momentum tensor on the FVD boundary and $S = S_{\mu \nu} h^{\mu \nu}$. 
In the thin-wall approximation, we can take $S_{\mu \nu} = -\sigma h_{\mu \nu}$, where $\sigma$ is the surface energy for the FVD boundary given in Eq.\,\eqref{eq:surface}. 
The $(t, t)$ and $(\theta, \theta)$ components of the first junction condition in Eq.\,\eqref{eq:1st_Israel} give
\begin{align}
a^2  = a_{1}^2 = a_{2}^2 \,.
\end{align}
On the other hand, the $(t, t)$ and $(\theta, \theta)$ components in Eq.\,\eqref{eq:2nd_Israel} give
\begin{align}
\label{eq:2nd_Israel_sigma}
[\xi^{\mu} \partial_{\mu} \ln (a_{2} \chi)]_{0} = - \frac{\sigma}{2M_{p \ell}^2} \,, 
\quad
[\xi_{\mu} D_{\tau} v^{\mu}]_{0} = - \frac{\sigma}{2M_{p \ell}^2} \,,
\end{align}
with $D_{\tau} v^{\mu} = \partial_{\tau} v^{\mu} + \Gamma^{\mu}_{\alpha \beta} v^{\alpha} v^{\beta}$, where $\Gamma^{\mu}_{\alpha \beta}$ is the Christoffel symbol defined in the FLRW metric. 

The EoM for the FVD boundary is given by~\cite{Israel:1966rt}
\begin{align}
\label{eq:EoM_Israel}
\bar{K}_{\mu \nu} S^{\mu \nu} = \left[ T_{\mu \nu} \xi^{\mu} \xi^{\nu} \right]_{0} \,.
\end{align}
If the energy-momentum tensor takes the perfect fluid form $T_{\mu \nu} = (\rho + p) u_{\mu} u_{\nu} + p g_{\mu \nu}$ with $u_{\mu}$ is the fluid 4-velocity, the EoM in Eq.\,\eqref{eq:EoM_Israel} can be expressed as
\begin{align}
\label{eq:EoM_Israel_rhop}
\left\{ \xi_{\mu} D_{\tau} v^{\mu} + 2 \xi^{\mu} \partial_{\mu} \ln (a_{2} \chi) \right\}_{0} = - \frac{2}{\sigma} \left[ (\rho + p) (u_{\mu} \xi^{\mu})^2 + p \right]_{0} \,.
\end{align}
In our setup, the energy density $\rho$ is composed of the radiation component $\rho_{R}$ and the vacuum energy component $\rho_{V}$. 

The shell energy conservation is given by 
\begin{align}
v^\nu D_\mu S^{\mu}_\nu  = \left[ T_{\mu \nu} \xi^{\mu} v^\nu \right]_{0} \,.    
\end{align}
Since $S_{\mu\nu}=-\sigma h_{\mu\nu}$, we obtain 
\begin{align}
0 = \left[ (\rho+p) u_\mu \xi^{\mu} u_\nu v^\nu \right]_{0} \,.    
\end{align}

We here assume the following condition:
\begin{align}
\label{eq:fluid_assumption}
[u^{\mu} \xi_{\mu}]_{0} =0 \,.
\end{align}
This condition implies that the radiation fluid flows smoothly through the FVD boundary, namely, there is no friction between the shell and the fluid. 
Then, from the shell energy conservation, we obtain 
\begin{align}
\label{eq:density_continuity}
\quad [\rho_{R}]_{0} = 0 \,, 
\end{align}
which implies 
\begin{align}
\label{eq:density_discontinuity}
\quad [\rho]_{0} = -\rho_V \,,
\end{align}
in our setup. 

Combining Eqs.\,\eqref{eq:2nd_Israel_sigma}, \eqref{eq:EoM_Israel_rhop} and \eqref{eq:fluid_assumption}, the EoM for the FVD boundary can be expressed by 
\begin{align}
\ddot{\chi} + (4 - 3a^2 \dot{\chi}^2) H \dot{r} + \frac{2}{a^2 \chi} (1 - a^2 \dot{\chi}^2) = \left( \frac{3\sigma}{4M_{p\ell}^2} - \frac{\rho_{V}}{\sigma}  \right) \frac{(1 - a^2 \dot{\chi}^2)^{3/2}}{a} \,,
\end{align}
which gives Eq.\,\eqref{eq:EOM} in the main text.

\bibliographystyle{JHEP}
\bibliography{reference}

\end{document}